\documentclass[conference]{IEEEtran}
\IEEEoverridecommandlockouts
\usepackage{cite}
\usepackage{amsmath,amssymb,amsfonts}
\usepackage{algorithmic}
\usepackage{graphicx}
\usepackage{textcomp}
\usepackage{xcolor}
\def\BibTeX{{\rm B\kern-.05em{\sc i\kern-.025em b}\kern-.08em
    T\kern-.1667em\lower.7ex\hbox{E}\kern-.125emX}}
\begin{document}

\title{Deep COVID-19 Recognition using Chest X-ray Images: A Comparative Analysis}

\author{\IEEEauthorblockN{Selvarajah Thuseethan, Chathrie Wimalasooriya and Shanmuganathan Vasanthapriyan}
	\IEEEauthorblockA{\textit{Department of Computing and Information Systems} \\
		\textit{Sabaragamuwa University of Sri Lanka}\\
		Sri Lanka \\
		\{thuseethan, wimalasooriya, priyan\}@appsc.sab.ac.lk}
}

\maketitle

\begin{abstract}
The novel coronavirus variant, which is also widely known as COVID-19, is currently a common threat to all humans across the world. Effective recognition of COVID-19 using advanced machine learning methods is a timely need. Although many sophisticated approaches have been proposed in the recent past, they still struggle to achieve expected performances in recognizing COVID-19 using chest X-ray images. In addition, the majority of them are involved with the complex pre-processing task, which is often challenging and time-consuming. Meanwhile, deep networks are end-to-end and have shown promising results in image-based recognition tasks during the last decade. Hence, in this work, some widely used state-of-the-art deep networks are evaluated for COVID-19 recognition with chest X-ray images. All the deep networks are evaluated on a publicly available chest X-ray image datasets. The evaluation results show that the deep networks can effectively recognize COVID-19 from chest X-ray images. Further, the comparison results reveal that the EfficientNetB7 network outperformed other existing state-of-the-art techniques.
\end{abstract}


\section{Introduction}
\label{sec:introduction}
Coronaviruses are a group of related viruses that can cause respiratory tract infections, ranging from mild to lethal. In 2019, a novel coronavirus called Severe Acute Respiratory Syndrome Coronavirus 2 (SARS-CoV-2) emerged from Wuhan, China and infected more than 229 million people with a mortality of over 4.7 million\footnote{https://covid19.who.int/}. The SARS-CoV-2 cause respiratory disease, namely coronavirus disease 19 (COVID-19), which is the key reason for the current COVID-19 pandemic \cite{velavan2020covid}. This virus can spread from one person to another primarily through the droplets of an infected person. Since the first major outbreak different variants of SARS-CoV-2 have been identified, which makes the journey towards controlling the pandemic more challenging. 

Even the most developed countries like the United States of America, the United Kingdom, Italy, Australia and Germany are heavily affected by COVID-19. The COVID-19 pandemic has changed the world's economy upside down, putting most of the countries in recession \cite{accikgoz2020early}. In addition, COVID-19 profound various societal challenges, causing serious psychological disorders like depression. The consequences of COVID-19 pandemic, such as increased suicidal rates, are expected to stay longer than the actual pandemic period \cite{sher2020impact}. Hence, in order to save the world population from these existing and future challenges, it is important that the COVID-19 has to be effectively handled.

Despite the efforts made towards inventing vaccines to control SARS-CoV-2, several computed aided techniques have been proposed for modelling and forecasting the spread pattern \cite{fanelli2020analysis}. Further, the machine learning-based COVID-19 recognition techniques play a vital role in the timely diagnosis of patients. This helps to provide prompt medication and largely prevent the virus from the spread. The majority of the existing COVID-19 recognition algorithms used multimodal inputs \cite{fang2021deep}, pre-processing \cite{zhang2020automated} and complex models \cite{hou2021automatic} to obtain benchmark performances. The state-of-the-art deep networks are however well known for image classification tasks with consistent results. More importantly, deep networks are capable to be trained end-to-end, where the input data can be given naturally. Hence, in this work, a set of popular deep networks are trained to effectively detect COVID-19 from chest X-ray images. The key contributions of this paper are as follows.
\begin{itemize}
	\item The state-of-the-art deep networks, such as VGG16, VGG19, DenseNet121, DenseNet201, InceptionV3, ResNet101, ResNet152, Xception, EfficientNetB0, EfficientNetB7, NASNetLarge, NASNetMobile, MobileNetV2, MobileNetV3 Small and MobileNetV3 Large are trained on a chest X-ray dataset for effective recognition of COVID-19 disease.
	
	\item A summary of the comparative results is analysed to evaluate the feasibility of each deep learning model.
	
\end{itemize}

\begin{figure*}[h]
	\begin{center}
		\includegraphics[width=\textwidth]{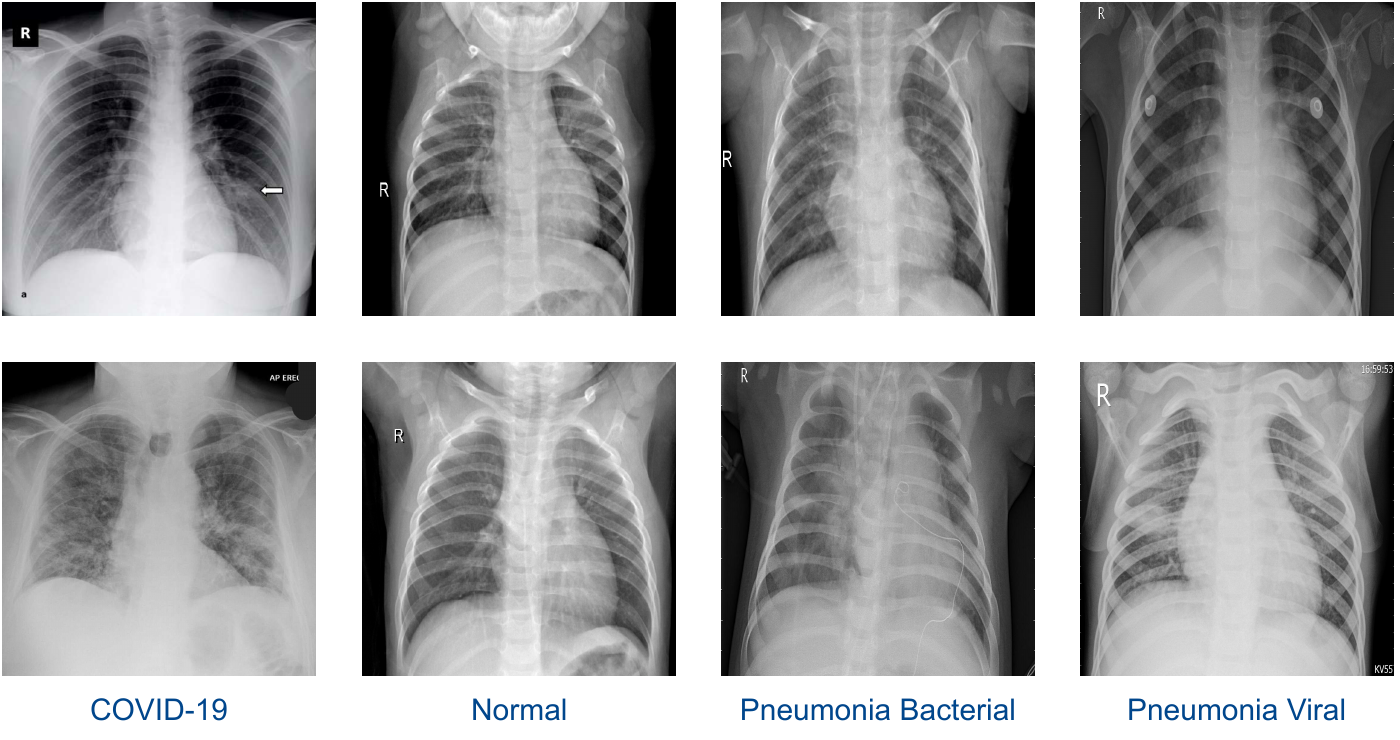}
	\end{center}
	\caption{The standard training and testing procedure employed.}
	\label{fig:dataset_samples}
\end{figure*}

The remaining sections of the paper are structures into three sections.  
The remainder of this paper is organized into four sections. A set of literature for COVID-19 recognition via chest X-ray images is comprehensively reviewed in \textit{Section \ref{sec:relatedwork}}. In \textit{Section \ref{sec:comparativeanalysis}}, the results obtained for all of the state-of-the-art deep networks are compared and analysed. This conclusion of the paper is presented in \textit{Section \ref{sec:conclusion}}.

\section{Related Work}
\label{sec:relatedwork}
Deep learning-based techniques have significantly improved the performance of various image classification tasks, ranging from emotion recognition \cite{wu2018adaptive} to agriculture \cite{afifi2021convolutional}. In the last two years, multiple approaches have been proposed for COVID-19 recognition. Some of the closely related works are reviewed in this section. However, authors are directed to read the surveys on the machine and deep learning-based COVID-19 infection recognition provided in \cite{rehman2021covid} and \cite{shoeibi2020automated}.

In one of the early works, a novel deep convolutional neural network (CNN) model called CoroNet is proposed for automatic detection of COVID-19 infection with chest X-ray images \cite{khan2020coronet}. The experimental results proved that the proposed CoroNet model achieved promising results even in small datasets. Authors also claim that the performance of the CoroNet can further be improved with additional training data. In \cite{guellil2020web}, a deep learning-based framework (COVIDz) is proposed to systematically predict the existence of COVID-19. The presented COVIDz framework yielded a classification accuracy of 99.64\% and an F-score of 99.20\%. Noor and Kareem used a deep CNN to build a COVID-19 diagnosis approach that showed over 94\% classification accuracies in three different tasks \cite{qaqos2020covid}. Meanwhile, the attention-based deep neural networks also attained benchmark performances in COVID-19 detection from X-ray images \cite{sitaula2021attention}.

In another recent work, a deep CNN based architecture is proposed for automated COVID-19 detection \cite{mostafa2021covid}. The abnormalities of the samples like low-resolution images are enhanced by a well-structured data augmentation technique. In \cite{bhatia2021transfer}, a transfer learning mechanism is adapted to implement the proposed approach. Ahsan et. al. \cite{ahsan2021detecting} designed a COVID-19 detection technique utilizing a deep CNN. In their study, various deep networks are considered as the backbone. A machine learning-based analytical framework also demonstrated better results in COVID-19 detection \cite{johri2021novel}. A deep CNN model with chest X-ray images taken from portable devices is used to effectively detect the COVID-19 \cite{de2020deep}. In this work, the characteristics that are common to COVID-19 and other diseases are differentiated using a joint response approach. An integrated stacking InstaCovNet-19 with multiple pre-processing techniques are presented to perform COVID-19 patient classification \cite{gupta2021instacovnet}.

According to the literature, most of the existing works use multiple model data, complex deep network architecture or heavy pre-processing. Motivated by this, a comparative analysis of the state-of-the-art for COVID-19 recognition is conducted and discussed in the next section.

\begin{figure*}[h]
	\begin{center}
		\includegraphics[width=\textwidth]{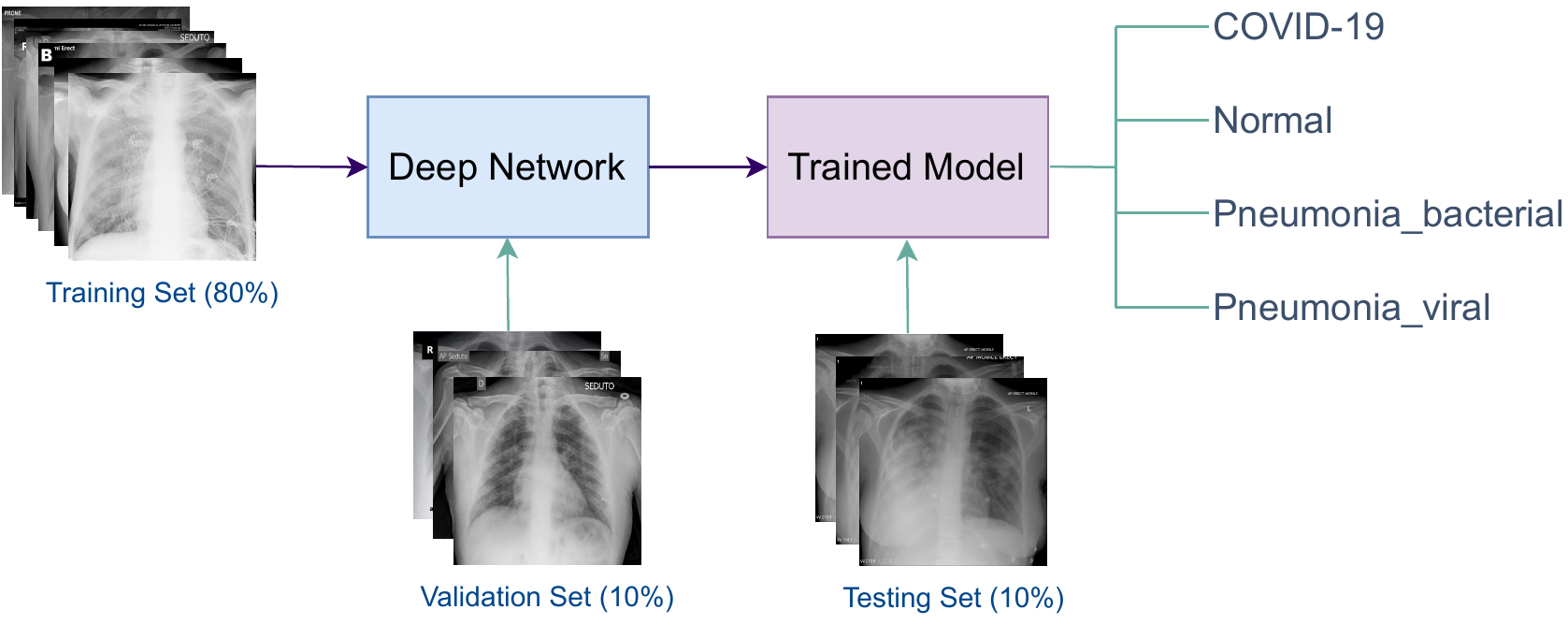}
	\end{center}
	\caption{The standard training and testing procedure employed.}
	\label{fig:training_procedure}
\end{figure*}

\section{Comparative Analysis}
\label{sec:comparativeanalysis}
In this section, the details about the state-of-the-art deep networks compared, the dataset used, implementation information with standard training and testing procedure and the results are discussed.

\subsection{State-of-the-art Deep Networks}
Various state-of-the-art deep networks are evaluated in this study. The list of state-of-the-art deep networks that are compared in this study is given below.

\begin{enumerate}
	\item \textbf{VGG16} and \textbf{VGG19} are the earliest and simplest CNN architectures, significantly increased the depth with very small receptive fields \cite{simonyan2014very}. A set of convolutional layers followed by three fully connected layers are stacked to construct both of these deep architectures.
	
	\item \textbf{DenseNet121} and \textbf{DenseNet201} have introduced a dense connection mechanism, where each layer of the network is connected to every other layer \cite{huang2017densely}. In addition to strengthening the feature propagation, reducing the number of training parameters and enhancing the feature reuse, both of these networks also alleviate the gradient vanishing problem.
	
	\item \textbf{InceptionV3} architecture is constructed based on the ideas of multiple researchers developed over the years \cite{szegedy2016rethinking}. Factorized convolutions and intensive use of regularization are the keys of the InceptionV3 in effectively handling the added computational complexity.
	
	\item \textbf{ResNet101} and \textbf{ResNet152} are capable of attaining higher accuracy and easier to optimize for image recognition tasks, in comparison to existing state-of-the-art deep networks \cite{he2016deep}. A residual learning process along with appropriately placed shortcut connections are introduced in these networks.
	
	\item \textbf{Xception} was built based on depthwise separable convolution concept \cite{chollet2017xception}. The depth multiplier is set to 1 for all the depthwise separable convolution layers. Throughout the architecture, a batch normalization layer is placed after each traditional convolution and depthwise separable convolution layer.
	
	\item \textbf{EfficientNetB0} and \textbf{EfficientNetB7} are two prominent network variants in the EfficientNets family of models. A scaling mechanism proposed in \cite{tan2019efficientnet} uses a compound coefficient to uniformly scale the dimensions, such as depth, width and resolution with a fixed ratio.
	
	\item \textbf{NASNetLarge} and \textbf{NASNetMobile} are two NASNet variations that are primarily designed for general and mobile deployment purposes, respectively \cite{zoph2018learning}. To design the NASNet models, the best convolutional layers with their parameter settings are realized through a novel search space. In addition, a new regularization scheme, namely ScheduledDropPath is also integrated to improve the generalization capability.
	
	\item \textbf{MobileNetV2}, \textbf{MobileNetV3 Small} and \textbf{MobileNetV3 Large} are based on both depthwise separable convolutions and pointwise convolutions \cite{howard2017mobilenets}. In MobileNets, for each input channel, single depthwise convolution filter is applied before combining the output using pointwise convolution.
\end{enumerate}

\subsection{Chest X-ray Dataset}
The chest X-ray image dataset\footnote{https://github.com/drkhan107/CoroNet} used in extensive experiments contains four classes, such as COVID-19, Normal, Pneumonia Bacteria and Pneumonia Viral. Figure \ref{fig:dataset_samples} shows two samples for each class available in the dataset. The publicly available version of the dataset consists of $320$, $445$, $449$ and $424$ image samples for COVID-19, Normal, Pneumonia Bacteria and Pneumonia Viral classes, respectively. The images available within this dataset are in different resolutions, which are later resized to the same dimension during the training process in supporting the deep networks individually. For instance, images in the dataset are resized to $224 \times 224$ in order to match the input image size of the VGG16 network.

\begin{figure*}
	\begin{center}
		\includegraphics[width=\textwidth]{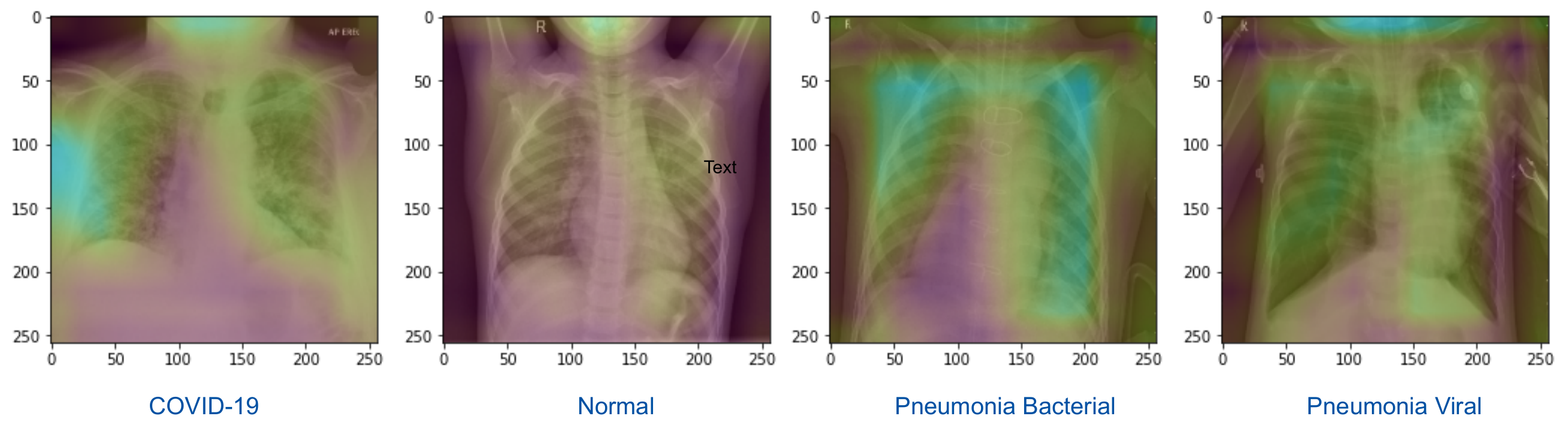}
	\end{center}
	\caption{The visual illustration of the spatial support obtained for each class by the EfficientNetB7 model.}
	\label{fig:saliency}
\end{figure*}

\subsection{Implementation, Training and Testing}
All the networks are implemented using the Tensorflow\footnote{https://www.tensorflow.org/} platform and evaluated on Google Colab. The deep networks are used as defined in TensorFlow Core v2.6.0. A traditional training and testing procedure is adapted, where 80\%, 10\% and 10\% samples of the whole dataset are used to train, validate and test the networks, respectively. Figure \ref{fig:training_procedure} shows the overall training and testing procedure. A stochastic gradient descent (SGD) optimizer is used with a momentum of 0.9, an initial learning rate of 0.05 and a decay of 0.0001.

\begin{table}[h]
	\caption{
		Comparison of state-of-the-art deep networks. Three best-performing models are highlighted.
	}
	\begin{center}
		\begin{tabular}{p{4cm}p{1cm}p{0.9cm}}
			\hline\hline\noalign{\smallskip}
			\textbf{Deep Network} 				& \textbf{Accuracy} 	& \textbf{Rank} 	\\
			\hline\noalign{\smallskip}
			
			VGG16			 					& 0.8040 		& 11 			\\
			VGG19								& 0.8162 		& 9 			\\
			
			DenseNet121	 						& 0.8639 		& 7 			\\
			DenseNet201	 						& 0.8792 		& 5 			\\
			
			InceptionV3							& 0.8863 		& 4 			\\
			
			\textbf{ResNet101}				 	& \textbf{0.9037} 		& \textbf{3} 			\\
			\textbf{ResNet152}				 	& \textbf{0.9138} 		& \textbf{2} 			\\
			
			Xception						 	& 0.8215 		& 8 			\\
			
			EfficientNetB0					 	& 0.8769 		& 6 			\\
			\textbf{EfficientNetB7}				& \textbf{0.9554}		& \textbf{1} 			\\
			
			NASNetLarge						 	& 0.7608 		& 14 			\\
			NASNetMobile						& 0.7215 		& 15 			\\
			
			MobileNetV2						 	& 0.8147 		& 10 			\\
			MobileNetV3 Small					& 0.7628 		& 12 			\\
			MobileNetV3 Large					& 0.7635 		& 13 			\\
			\hline\hline\noalign{\smallskip}
		\end{tabular}
		\label{tab:results}
	\end{center}
\end{table}

\subsection{Results and Discussion}
The classification accuracy obtained for fifteen state-of-the-art deep networks are compared in this research. Table \ref{tab:results} illustrates the comparison results, where the deep networks that showed the best performances are highlighted. As can be seen, the EfficientNetB7 achieved the best results with 95.54\% of classification accuracy. The second and third best performances were shown by both ResNet variants, obtaining slightly more than 90\% classification accuracies. The lightweight networks, such as NASNetMobile, MobileNetV2, MobileNetV3 Small and MobileNetV3 Large also showed comparable performance in recognizing COVID-19. Among them, the MobileNetV2 yielded the best accuracy with 81.47\%, followed by MobileNetV3 Large, MobileNetV3 Small and NASNetMobile. It is interesting to note that the very deep networks like InceptionV3 and DenseNet201 performed comparably low in COVID-19 recognition, which could be because of the lack of training data.

Figure \ref{fig:saliency} illustrates the saliency maps acquired by the best performing EfficientNetB7 model for one sample taken from COVID-19, normal, pneumonia bacterial and pneumonia viral classes. Note that the selected images are picked from the set of correctly classified samples. As can be seen, the corresponding spatial support attained for each class are clearly distinguishable.

\begin{figure}[h]
	\begin{center}
		\includegraphics[width=9cm]{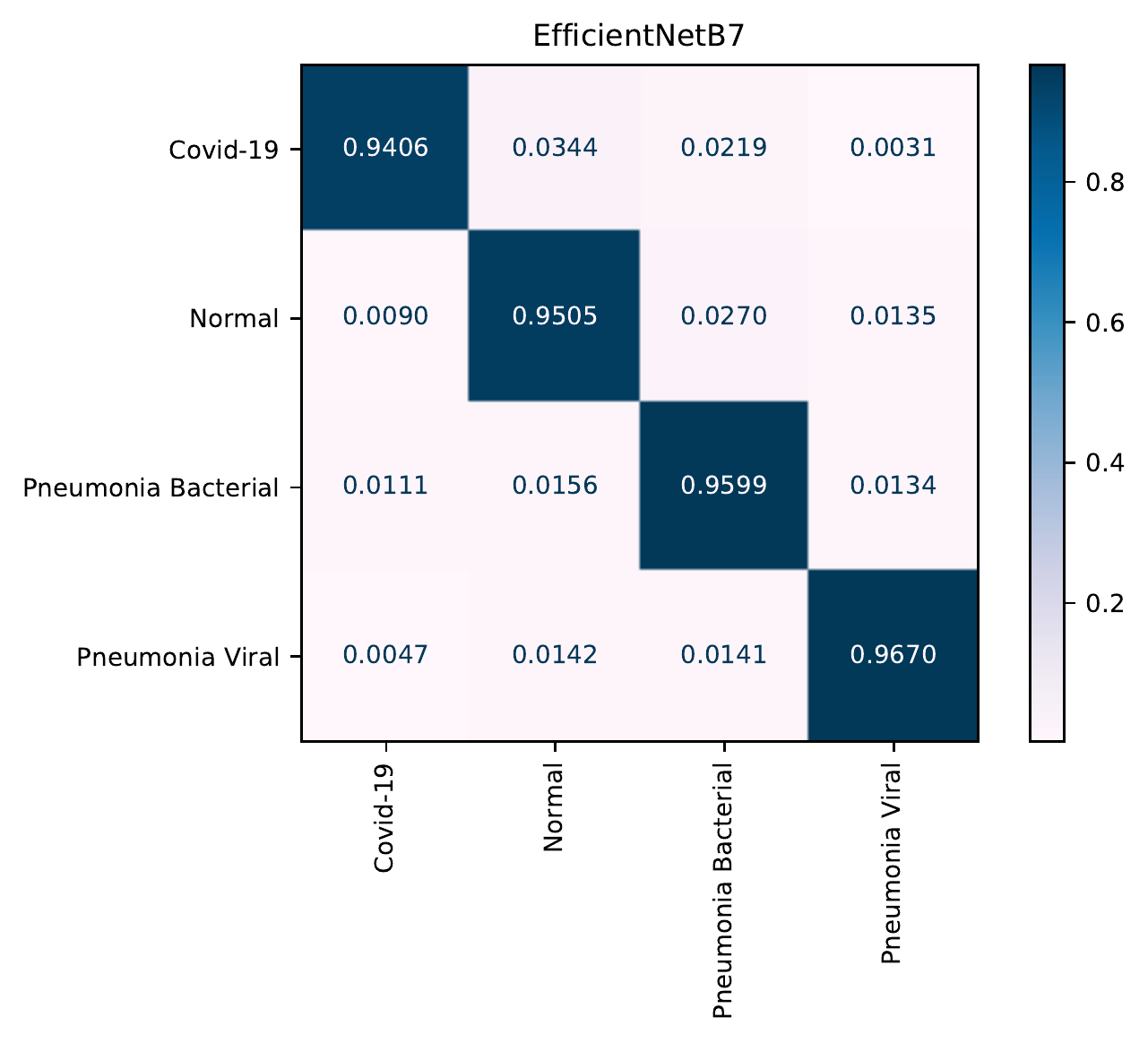}
	\end{center}
	\caption{The confusion matrix obtained for the best performed EfficientNetB7 model.}
	\label{fig:EfficientNetB7}
\end{figure}

The confusion matrix generated for the EfficientNetB7 deep network is presented in Figure \ref{fig:EfficientNetB7}. The COVID-19 class achieved a slightly lower accuracy of 94.06\%, in comparison to other classes. The majority form the misclassified COVID-19 samples are confused with the normal class. Amongst all other classes, Pneumonia Viral achieved the best accuracy of 96.70\%. The Normal and Pneumonia Bacterial classes also reached more than 95\% of accuracy. Overall, the performance manifested all of the classes are at the expected level.

\section{Conclusion}
\label{sec:conclusion}
The global outbreak of COVID-19 has impacted almost every country and its people. Hence, it is vital to devise an effective and timely way of recognizing COVID-19. Adapting the superiority of deep learning techniques, many approaches have been proposed in the past. Many of them are limited due to integrated pre-processing, complex structures and the use of multiple model data. Knowing the fact that the existing deep networks are end-to-end trainable and powerful in extracting visual features, in this paper, a comprehensive comparative analysis is performed. The results obtained for fifteen widely-used deep networks are compared and analysed. Comparison results reveal that the EfficientNetB7 network demonstrated the best classification accuracy in recognizing COVID-19 with chest X-ray images. Analysing other COVID-19 datasets could be a potential future work of this research.

\vspace{12pt}

\end{document}